\documentclass[onecolumn, 12pt, secnumarabic,amssymb,nobibnotes,showpacs,superscriptaddress,longbibliography]{revtex4-1}[]
\usepackage[english]{babel}
\usepackage{graphicx}
\usepackage{amsmath}
\usepackage{amsfonts}
\usepackage[usenames,dvipsnames]{color}
\usepackage{hyperref}
\usepackage{blindtext}
\hypersetup{colorlinks}
\hypersetup{linkcolor=black, citecolor=black, urlcolor=blue}
\usepackage[utf8]{inputenc}
\usepackage{color}
\usepackage{titlesec}
\titleformat{\section}
  {\normalfont\fontsize{16}{15}\bfseries}{\thesection}{1em}{}
\titleformat{\subsection}[runin]{}{}{}{}[]
 \newcommand\mycolor{\color{black}\xspace}
\usepackage{setspace}
\usepackage{caption}
\begin{document}
\title{Complex electric double layers in charged topological colloids}
\author{Jeffrey C. Everts}
\email{jeffrey.everts@gmail.com}

\address{Department of Physics, Faculty of Mathematics and Physics, University of Ljubljana, Jadranska 19, 1000 Ljubljana, Slovenia}

\author{Miha Ravnik}
\address{Department of Physics, Faculty of Mathematics and Physics, University of Ljubljana, Jadranska 19, 1000 Ljubljana, Slovenia}
\address{Department of Condensed Matter Physics, Jozef Stefan Institute, Jamova 39, 1000
Ljubljana, Slovenia}

\date{\today}

\begin{abstract}
{\bf Charged surfaces in contact with liquids containing ions are accompanied in equilibrium by an electric double layer consisting of a layer of electric charge on the surface that is screened by a diffuse ion cloud in the bulk fluid. This screening cloud determines not only the interactions between charged colloidal particles or polyelectrolytes and their self-assembly into ordered structures, but it is also pivotal in understanding energy storage devices, such as electrochemical cells and supercapacitors. However, little is known to what spatial complexity the electric double layers can be designed. Here, we show that electric double layers of non-trivial topology and geometry -including tori, multi-tori and knots- can be realised in charged topological colloidal particles, using numerical modelling{\mycolor within a mean-field Poisson-Boltzmann theory}. We show that the complexity of double layers -including geometry and topology- can be tuned by changing the Debye screening length of the medium, or by changing the shape and topology of the (colloidal) particle. More generally, this work is an attempt to introduce concepts of topology in the field of charged colloids, which could lead to novel exciting material design paradigms.}
\end{abstract}

\maketitle

%Topology is an important concept in contemporary physics with many applications ranging from string theory \cite{Strominger:1985}, light fields \cite{Dennis:2010}, gravitational radiation \cite{Thompson:2014} and quantum computing \cite{Bonesteel:2005}, to the study of topological defects in magnets \cite{yu:2010}, liquid crystals \cite{Senyuk:2013}, superconductors \cite{Abrikosov:1957} and cosmology \cite{Kibble:1976}. It aided in the discovery of topological insulators where the topology of the band structure plays a paramount role \cite{Zhang:2011}. Recently, topological concepts are starting to pave their way into soft condensed matter physics, by the realization of a mechanical topological insulator \cite{Kane:2014}, the design of topologically protected transport of colloidal particles \cite{Loehr:2016}, and understanding the self-assembly of (uncharged) colloidal particles driven by the topological defects in a liquid crystalline host \cite{Tkalec:2011}, amongst others, leading to a range of special topology-conditioned material properties.
The interplay between topology and geometry is an important feature in contemporary physics, with many applications ranging from string theory \cite{Strominger:1985} on the smallest scales to the study of cosmology on the largest scales \cite{Kibble:1976}. Topology, for example, aided in the discovery of topological insulators where the topology of the band structure plays a paramount role \cite{Zhang:2011}. Recently, topological concepts are starting to pave their way into soft condensed matter physics, by e.g. the realization of a mechanical topological insulator \cite{Kane:2014}, the design of topologically protected transport of colloidal particles \cite{Loehr:2016}, and understanding the self-assembly of (uncharged) colloidal particles driven by the topological defects in a liquid crystalline host \cite{Tkalec:2011,Senyuk:2013}, leading to a range of topology-conditioned material properties. The study of particle geometry and surface functionality in soft matter, on the other hand, has been made possible by the tremendous advances in chemical synthesis \cite{Glotzer:2007}, which has been shown to affect the phase behaviour \cite{Glotzer:2012, Gantapara:2013} and swimming behaviour of complex-shaped particles \cite{Bechinger:2013}, but also how topological defects are coupled to the particle topology and geometry \cite{Martinez:2014}.

While charged systems are investigated thoroughly with topological concepts in hard condensed matter, surprisingly enough this is not the case in soft condensed matter where topology is only explored in the context of e.g., liquid crystals \cite{Musevic:2006,Smalyukh:2009,Machon:2013}, metamaterials \cite{Chen:2016}, membranes \cite{Lubensky:1991}, microemulsions \cite{Holyst:1997} and DNA \cite{Arsuaga:2005}.  Charged systems in soft matter emerge primarily due to ionic degrees of freedom rather than the electronic ones in their ``hard" counterpart. Specifically, charged surfaces that are formed in a liquid medium have a surface charge that stems from cationic and/or anionic surface groups \cite{Ninham:1971}, that are screened by the counter and/or co-ions in the medium that form a diffuse ion cloud with a decreasing net charge density in the bulk fluid \cite{Andelman}. Such a surface charge with its screening cloud - called the electric double layer -  has profound influence on macroscopic thermodynamic properties, which can be of relevance for supercapacitors \cite{Janssen:2017}, and further, profoundly affects the self-assembly of charged colloidal particles in bulk \cite{Fuji:1989, Leunissen:2005}, in external fields \cite{Liu:2014} and at air-water interfaces \cite{Wickman:1998}.

In this article we demonstrate how electric double layers of complex geometry and topology can be created, and how this couples to a non-trivial particle topology, unlike the topologically simple (charged) shapes that are known in the literature, such as rods \cite{Lowen:1994}, spheres \cite{Alexander:1984}, or ellipsoids \cite{Alvarez:2010}, which have been studied extensively in complex environments \cite{Loverde:2007, Luijten:2014}. We first generalise the electric double layer of spherical particles to toroidal particles and, subsequently, we explore the double layer cloud around a trefoil knot, together with its surface charge distribution for a simple charge-regulation mechanism where the electrostatic potential is fixed at the boundary. Later, we generalize our results to electric double layers of nontrivial topological shapes realised by particles in the form of torus knots, and demonstrate the coupling of the double layer topology with the topology and geometry of the particle.  In view of experiments, colloidal particles  \cite{Martinez:2014} as well as macromolecules  \cite{Lukin:2005,Mallam:2006} of nontrivial topological shapes seem to be to the first candidates for experimental realisation of double layers with a nontrivial geometry and topology.  Finally, this work is an attempt to bring the concepts of topology to the field of charged and ionic fluids, possibly allowing for new materials and new topology-determined interactions.

%the class of torus knots which can be realized either as colloidal particles and/or (macro)molecules
\section*{Results}
\subsection*{\bf General double layers and definition of topological shape.} Our approach to determine the electric double layer in a distinct geometry is to use the mean-field non-linear Poisson-Boltzmann (PB) equation which relates the electrostatic potential $\psi({\bf r})=k_BT\phi({\bf r})/q_e$ to the cation $\rho_+({\bf r})$, and anion $\rho_-({\bf r})$ density profiles by using Boltzmann distributions, with $k_BT$ the thermal energy, $\phi({\bf r})$ the dimensionless electrostatic potential, and $q_e$ the elementary charge. The characteristic decay length for $\phi({\bf r})$ and $\rho_\pm({\bf r})$ in the linear regime, $|\phi({\bf r})|\ll1$, is the Debye (screening) length $\lambda_D$ which can be tuned by varying the bulk ion density $\rho_s=\rho_\pm(\infty)$, since $\lambda_D\propto\rho_s^{-1/2}$, and $\lambda_D$ is the key parameter that can be used to control the double layer complexity and effective topological shape. Note that the PB approach does not cover some detailed features of the electric double layer such as packing \cite{Hartel:2015, Perkin:2017} and correlation effects \cite{Podgornik:2013}, however, out core interest is in \emph{general} topological and geometric features of the electric double layer for which PB is qualitatively good and, moreover, can be related to a broad variety of systems and setups (see Methods).

The electric double layer consists of a layer of surface charge density $q_e\sigma({\bf r})$ and a diffuse ion cloud that screens this surface charge, which we describe by the volume net ion charge density $q_e\rho({\bf r}):=q_e[\rho_+({\bf r})-\rho_-({\bf r})]$. We assume a positively charged particle and constant-potential boundary conditions at the particle surface from which it follows that $\rho_0=\rho({\bf r}), \ {\bf r}\in\Gamma$, is constant at the particle surface $\Gamma$, which allows us to define a normalized net charge $\rho({\bf r})/\rho_0\in[0,1]$. Note that $\rho_0$ is not constant along $\Gamma$ for other boundary conditions. Finally, in Ref. \cite{Everts:2015} we showed that the constant-potential boundary condition approximates a charge regulation mechanism where both cations and anions can adsorb on the particle surface, so we made this choice for constant-potential particles not only for convenience, but it is also a realistic boundary condition. 

\begin{figure}[t]
\centering
\includegraphics[width=\textwidth]{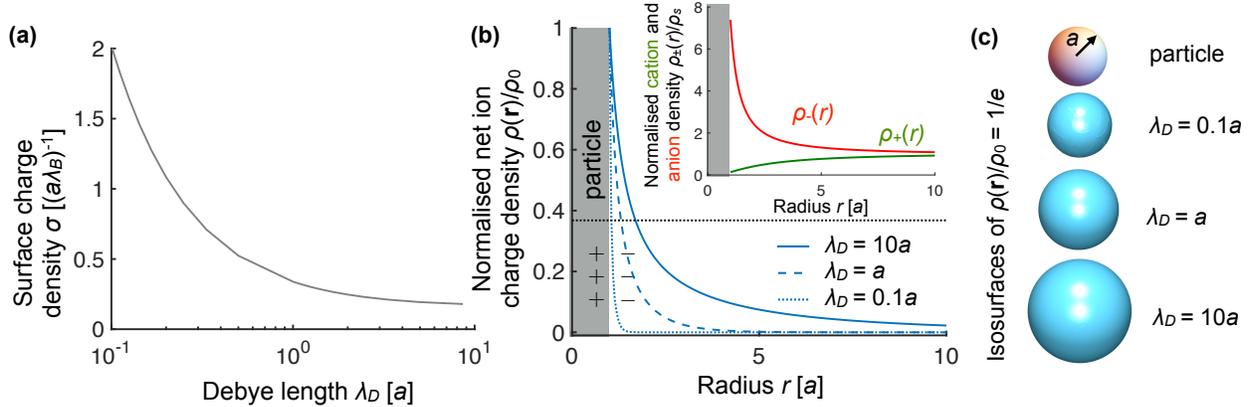}
\caption{\textbf{Electric double layer of a positively charged sphere.} We fix the radius $a$ and fix the dimensionless surface potential $\Phi_0=2$, and vary the Debye lengths $\lambda_D$. The double layer consists of a constant surface charge density $\sigma$ as depicted in (a), with $\lambda_B$ the Bjerrum length, and a diffuse ion cloud characterized by the local net charge density as defined by the difference of the cation and anion density, $\rho(r)=\rho_+(r)-\rho_-(r)$, see (b). Here $\rho_0=\rho(r=a)$, such that $\rho(r)/\rho_0$ varies between 0 and 1. In (c) we plot various isosurfaces of $\rho(r)$ defined by the set of points $\bf r$ in the system volume such that $\rho({\bf r})/\rho_0=1/e$, and see that regardless of the value for $\lambda_D$ all the isosurfaces are spheres.}
\label{fig:sphere}
\end{figure}

As an example, we show $\sigma({\bf r})$ and $\rho({\bf r})/\rho_0$ for a positively charged colloidal sphere with constant (dimensionless) surface potential $\Phi_0=2$ in Figure \ref{fig:sphere}. Since we assume the particle to be positively charged, the screening cloud is negatively charged, and the whole system is globally charge neutral. Due to the symmetry of the particle and the surrounding environment, the surface charge density $\sigma$ is constant along the particle surface and  it decreases with the increasing Debye screening length $\lambda_D$, which is equivalent to decreasing $\rho_s$, see Figure \ref{fig:sphere}(a). The discharging can be rationalized by the law of mass action: if less ions are available in the bulk, less ions will adsorb on an otherwise neutral surface in the case of charging by adsorption.

The net charge density of the negatively charged diffuse screening cloud around the positively charged sphere decays monotonically with the radial distance $r$ from the center of the particle, and is more diffuse with increasing Debye screening length (see Figure \ref{fig:sphere}(b)). Another way to visualize the screening cloud is by plotting isosurfaces of the normalized charge density $\rho({\bf r})/\rho_0$ in the system volume $V$, where a distinct isosurface is defined as $\Gamma_{C}=\{{\bf r}\in V: \rho({\bf r})/\rho_0=C\}$, with $C$ a constant. The set of all these isosurfaces uniquely describe the whole geometry and structure of the screening cloud, and interestingly, for a fixed $C$ these isosurfaces can act as topological objects, which we will now demonstrate. Note that a specific value of $C$ has no physical significance due to the diffuse nature of the double layer, but nevertheless the isosurfaces provide insights in the full three-dimensional shape of the screening cloud. Hence, without loss of generality, we choose the surface $\Gamma_{1/e}=\{{\bf r}\in V: \rho({\bf r})/\rho_0=1/e \ =0.367...\}$ - $e=2.718...$ is the Euler number -, the isosurface where the net ion density has decayed over one Debye length within the linear screening regime for a charged flat plate. When we investigate the topological shape of these isosurfaces for the spherical particle upon varying the Debye length $\lambda_D$, we see that independent of $\lambda_D$ the isosurface $\Gamma_{1/e}$ has the topological shape of a sphere and that these spherical isosurfaces grow with increasing $\lambda_D$. This generic behaviour does not change if we would take a different value for $C$. In the following we will define the double layer topology by the topology of such an isosurface and investigate its consequences for various particle shapes with non-trivial topology. 

\subsection*{\bf The electric double layer of a charged colloidal torus.}

 \begin{figure*}[t]
\centering
\includegraphics[width=0.95\textwidth]{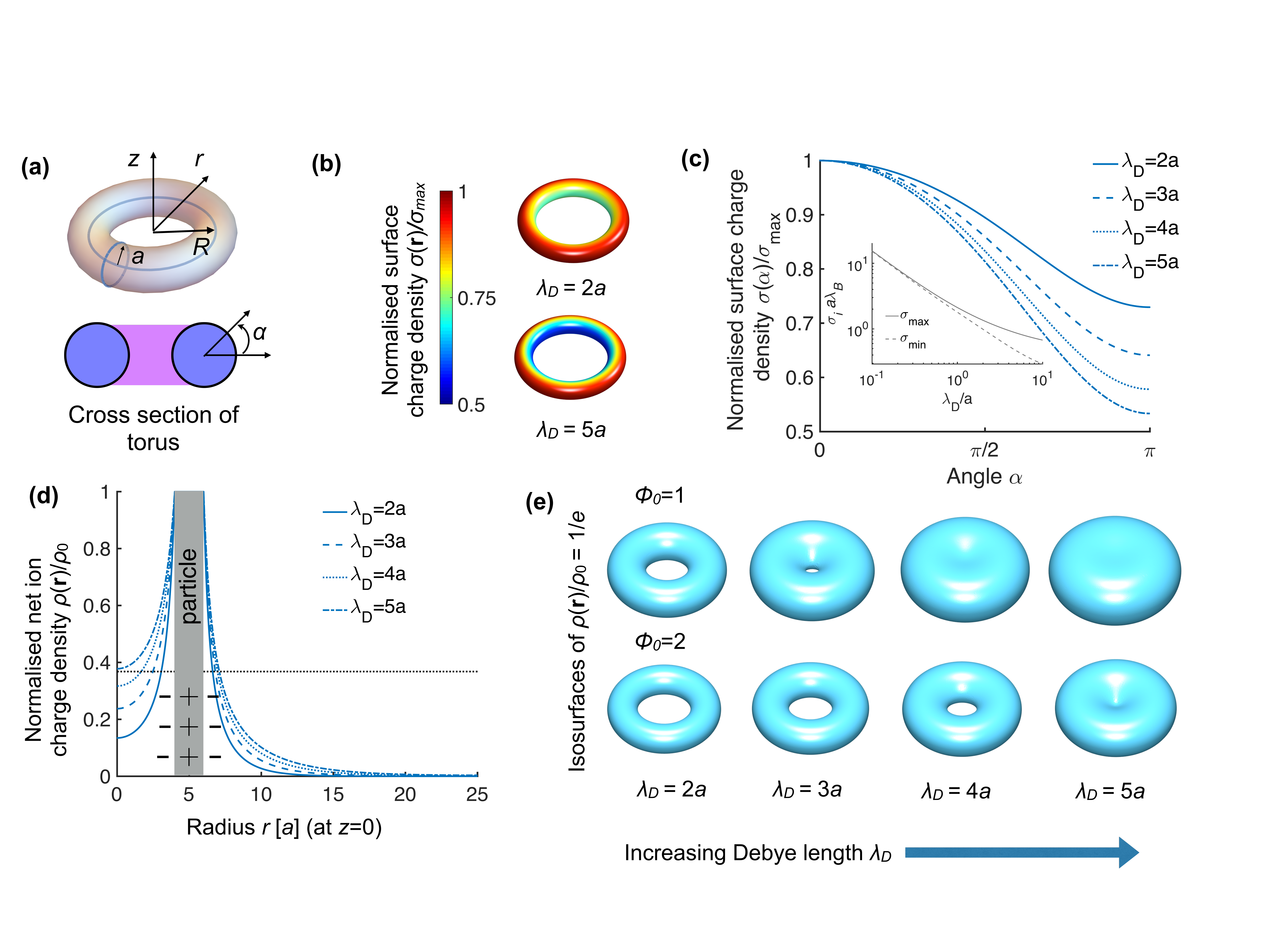}
\caption{\textbf{Electric double layer of a charged toroidal particle.} (a) Scheme of the torus with inner radius $a$ and outer radius $R=5a$ and constant (dimensionless) surface potential $\Phi_0=2$, with angle $\alpha$ defined. (b) Inhomogeneous surface charge density $\sigma({\bf r})$ at the surface of the torus for two representative values of the Debye length $\lambda_D$ with stronger inhomogeneities for larger $\lambda_D$. (c) Angular dependence of the full surface charge distribution $\sigma(\alpha)$ for some values of $\lambda_D$. In the inset we show the minimum (maximum) surface charge density $\sigma_\text{min}$ ($\sigma_\text{max}$) as function of $\lambda_D$ showing that inhomogeneities are smaller for small $\lambda_D$ since $\sigma_\text{max}$ converges to $\sigma_\text{min}$. (d) Net ion charge density of the diffuse ion cloud $\rho(r)/\rho_0$ along the radial coordinate $r$ at $z=0$. (e) Selected isosurfaces of the net ion charge density defined by $\rho({\bf r})/\rho_0=1/e$, which have the topological shape of a torus or a sphere depending on the value of $\Phi_0$ and $\lambda_D$. Note that the value of $\rho_0$ is different for the results of $\Phi_0=1$ and $\Phi_0=2$. }
\label{fig:torus}
\end{figure*}

The torus is a particle with non-trivial topology with the geometry shown in Figure \ref{fig:torus}(a). It is topologically distinct from a sphere since it is has a topological invariant called genus $g=1$, or stated differently, one hole in the surface (the sphere has $g=0$). When we investigate the electric double layer of such a particle, we see that unlike the sphere, the torus has an inhomogeneous charge distribution for constant-potential boundary conditions, see Figure \ref{fig:torus}(b) and (c). This is because the charged surface acts as an ideal conductor and therefore, effectively, the surface charges separate to be as far as possible from each other to reduce the electrostatic energy. This results in a higher charge accumulation at the outer rim of the torus, as is shown in Figure \ref{fig:torus}(b), with the explicit angular dependence in Figure 2(c). For small $\lambda_D$, we find that $\sigma({\bf r})$ becomes more homogeneous because of enhanced screening, but this is also accompanied by an increase in the total charge of the colloidal particle, see the inset in Figure \ref{fig:torus}(c) where the maximal value of the surface charge density $\sigma_\text{max}$ and the minimal value $\sigma_\text{min}$ start to coincide for small $\lambda_D$, and both quantities increase for decreasing $\lambda_D$.

The screening cloud on the torus -characterized by the net ion charge density $\rho({\bf r})$- has different topological shapes than the double layer of a sphere. In Figure \ref{fig:torus}(d) we show the full net ion charge density along the radial axis through the center of the torus. Note that we chose the particle to be positively charged and that the ion cloud bares a net negative charge (for negatively charged particles the results are equivalent, only the surface and ion densities change sign). The isosurface $\Gamma_{1/e}$ (see Figure \ref{fig:torus}(e)) has a toroidal topological shape for small $\lambda_D$ where the the isosurface ``inherits" the shape of the particle, and changes to a spherical topology by closing the center hole when $\lambda_D$ is sufficiently large. Also note that the value of the surface potential $\Phi_0$, here chosen to be well inside the non-linear regime of Poisson-Boltzmann theory, determines also for which $\lambda_D$ the toroidal topological shape transforms in that a spherical topology.  In the case of $\Phi_0=1$ this occurs between $\lambda_D=3a$ and $\lambda_D=4a$, while for $\Phi_0=2$ this occurs between $\lambda_D=4a$ and $\lambda_D=5a$. Note, however, while comparing the results for the two different values of $\Phi_0$, that also the normalisation constant $\rho_0$ is different for the two results. {\mycolor Finally, the transformation in the topological shape of the double layer has been inferred by visual inspection, as will be done in the remainder of the paper. To not miss any transition, we performed calculations with small increments in the Debye length and inspected the resulting isosurfaces. If the topologies were more complex -beyond this paper, as for example in some porous materials-, one could calculate topological invariants ascribed to the isosurface, such as the Euler characteristic, establishing a methodological algorithm for identifying the topology of the double layer isosurfaces.}

In view of generality, taking the isosurface $C=1/e$ was an arbitrary selection, however, would we have taken another isosurface (a different $C$), the same topological shapes would emerge but in a different range of Debye lengths. More specifically, depending on the value chosen for $C$, the transformation from toroidal to spherical shape occurs at a different value of $\lambda_D$;{\mycolor a lower value of $C$ means that the transition occurs at lower $\lambda_D$.} Alternatively, in the analysis, one can also fix $\lambda_D$ and vary $C$ to find the same shape transformation as the one shown in Figure \ref{fig:torus}(e). From a more general perspective, we show that the net ion charge density surrounding a torus changes its topological shape if we vary the Debye length (e.g. in experiments controlled by the salt concentration), which is an interesting result of tuning the shape of the electric double layers. {\mycolor Finally, we note that when charge correlation effects are included, the net ion charge density is not necessarily monotonous as function of the distance from the charged surface (see e.g. Ref. \cite{Netz:2000, Bazant:2011}).  In this case, the topological sequences can depend on the value of the cut-off $C$ and can even result in non-connected isosurfaces. However, in this paper we will only consider the mean-field result that is applicable to monovalent (and possibly divalent) ions.}

\subsection*{\bf The trefoil knot particle.}

\begin{figure*}[t]
\centering
\includegraphics[width=0.95\textwidth]{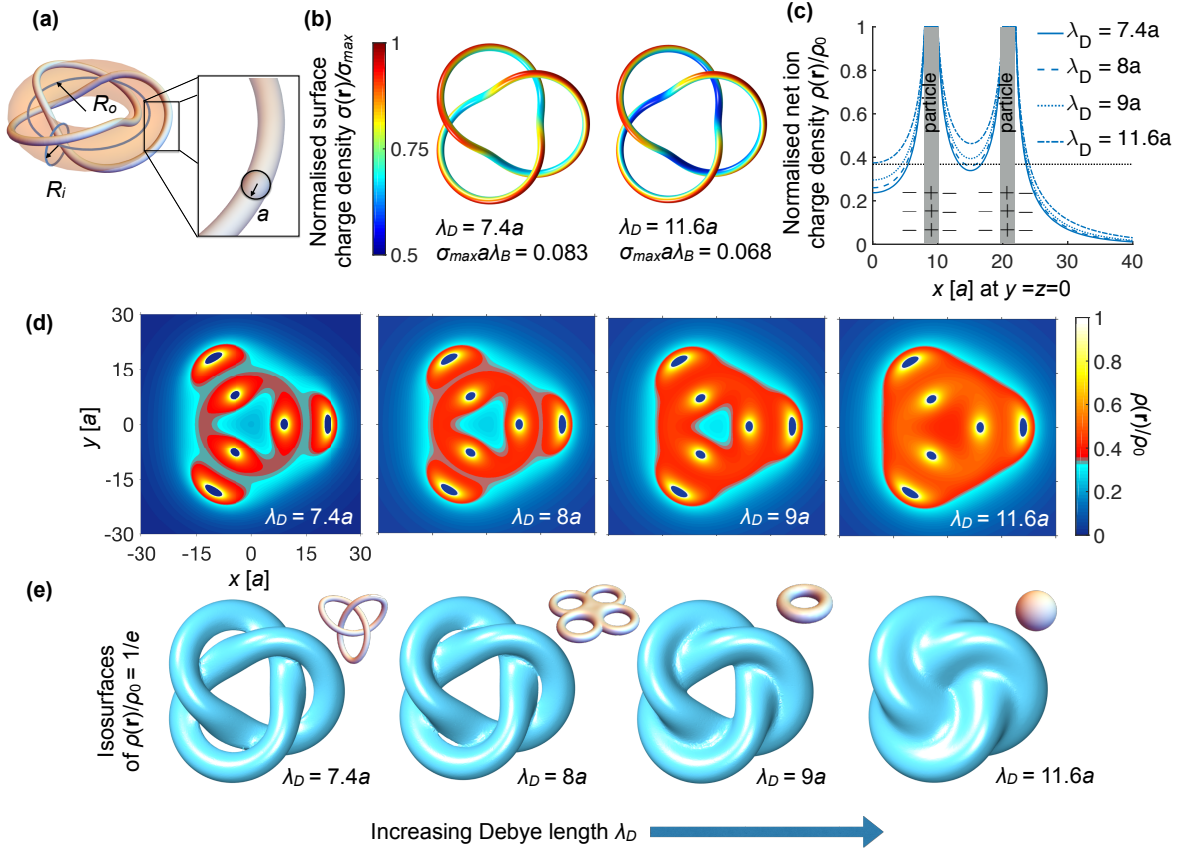}
\caption{\textbf{Electric double layer of a positively charged trefoil knot particle.} (a) The knot has tube radius $a$ and is wound around a torus with inner radius $R_i=6a$ and outer radius $R_o=15a$. We vary the Debye length $\lambda_D$ while keeping the constant dimensionless surface potential fixed at $\Phi_0=2$. In (b) we show with colormaps the surface charge density $\sigma({\bf r})$ normalized to its value $\sigma_\text{max}$. In (c) the net charge ion density of the negatively charged diffuse ion cloud is shown along the $x$-axis and in (d) in the $xy$-plane. The color scale in (d) is chosen such that the transition from blue to red is through the value $\rho({\bf r})/\rho_0=1/e$. (e) Isosurface of the net ion charge density in the screening cloud defined by $\rho({\bf r})/\rho_0=1/e$ for various Debye lengths $\lambda_D$. The double layer changes from a trefoil knot (first panel) to a genus-$4$ handlebody (second panel), to torus (third panel) and, finally, to a sphere (fourth panel). These isosurfaces are shown from various angles in Supplementary movie 1. The topologically equivalent shape of the isosurface is shown in each panel in the upper-right inset.}
\label{fig:trefoil}
\end{figure*}

We now consider a particle in the shape of a trefoil knot, which has a more complex double layer. A trefoil knot particle can be constructed from a line that lies on a surface of a (virtual) torus with inner radius $R_i$ and outer radius $R_o$, and by giving the line a finite tube radius $a$ using the parametrisation explained in the Methods, see Figure \ref{fig:trefoil}(a). In this case the trefoil knot is also a torus knot. We solve the PB equation for the trefoil knot particle and investigate the electric double layer for various $\lambda_D$. As for the torus, we find an inhomogeneous $\sigma({\bf r})$, which is conditioned by the geometry of the particle as shown in Fig. \ref{fig:trefoil}(b). The surface charge density profile is determined primarily by the interplay of two effects: (i) charges accumulate at the outer rim of the tube and (ii) the particle locally discharges if surfaces are close together, such as at particle tube crossings. Two representative surface charge density profiles are shown in Figure \ref{fig:trefoil} for two values of $\lambda_D$. Analogous to the torus, we see that the total charge on the particle decreases with increasing $\lambda_D$, which is accompanied by stronger inhomogeneities, as illustrated by the value of $\sigma_\text{max}$ and the colormap of $\sigma({\bf r})$, respectively.

In Figure \ref{fig:trefoil}(c) we plot the net charge density of the ion cloud surrounding the trefoil knot particle along the positive $x$-axis. The same density profiles occur along all axes in the $xy$-plane under $120^{\circ}$ with the $x$-axis because of the symmetry of the trefoil knot. Note the diffuse nature of the ion cloud and that various parts of the net charge density grow larger than the treshold $C=1/e$ (dotted black line in Figure \ref{fig:trefoil}(c)) depending on the value of $\lambda_D$. To better visualize the resulting double layer overlaps, we plot $\rho({\bf r})$ in the $xy$ plane in Figure \ref{fig:trefoil}(d), where $\rho({\bf r})/\rho_0>1/e$ is shown in red and  $\rho({\bf r})/\rho_0<1/e$ is shown in blue. We see that various types of double layer overlaps occur depending on the value of $\lambda_D$. 

Trefoil knot particles exhibit an even richer variation of topological shapes of the electric double layer as the torus, which is seen from the isosurfaces of the net ion charge density (again without loss of generality, we choose the isosurfaces $\Gamma_{1/e}$). In the high screening regime where $\lambda_D$ is small, the topology of $\Gamma_{1/e}$ is the same as that of the particle (first panel in Figure \ref{fig:trefoil}(e)), with the topological equivalent shape of $\Gamma_{1/e}$ shown in the upper-right inset. When we increase $\lambda_D$, double layer overlaps occur at the three crossing points of the trefoil knot and $\Gamma_{1/e}$ has the topology of a $4-$torus, i.e. a torus or handlebody with $g=4$. Upon further increase of the Debye length, double-layer overlaps occur at the three outer holes of the trefoil knot, and the topology of $\Gamma_{1/e}$ changes to that of a torus with $g=1$. Ultimately, the center hole closes and the double layer of the trefoil particle has the topological shape of the sphere. The isosurfaces from Figure \ref{fig:trefoil}(e) are also shown from various camera angles in Supplementary movie 1.

Hence, we demonstrated for the trefoil knot particle that there is a non-trivial shape transformation of $\Gamma_{1/e}$ by tuning $\lambda_D$. The resulting topological shapes are not necessarily that one of the sphere or the one of the particle. Again in view of generality, the same transformations between topological shapes can be found with another value for $C$, albeit the range of $\lambda_D$ to see all shapes is different than for the example shown here. 

\newpage
\subsection*{\bf Topological shape of the double layer for torus knot particles.}
\begin{figure}[ht]
\centering
\includegraphics[width=0.82\textwidth]{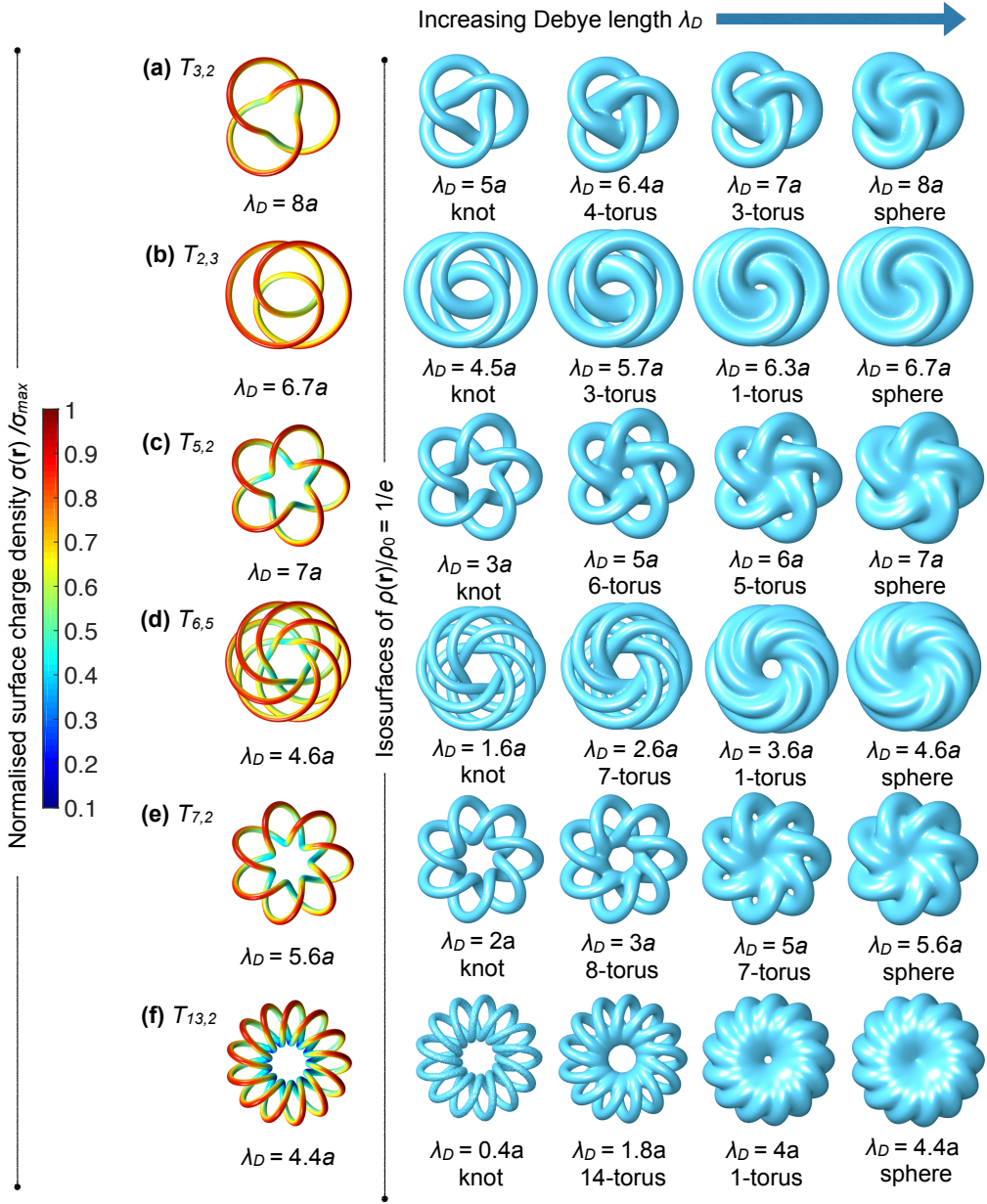}
\caption{\textbf{Electric double layer topological shapes for a selection of torus knots $T_{p,q}$.} The torus knot is wound around a torus with inner radius $R_i=6a$ and outer radius $R_o=12a$, with $a$ the tube radius, and we fix $\Phi_0=2$. In the first panel of each row, we show the surface charge density $\sigma({\bf r})$, normalized to the maximal value $\sigma_\text{max}$ for a representative value of the Debye length $\lambda_D$. In the remaining four columns we show the isosurfaces defined by $\rho({\bf r})/\rho_0=1/e$ and tune $\lambda_D$. These isosurfaces transform from a knot with genus $g=(p-1)(q-1)/2$ (first panel of each row), to a $(p+1)$-torus, to a $g$-torus with $g=1$ (b,d,f) or $g=p$ (a,c,e), and ultimately to a sphere. }
\label{fig:tknots}
\end{figure}

We have shown that different particle topologies (torus, trefoil knot) result in different isosurface topological shapes of the screening cloud depending on the value of $\lambda_D$. Since the double layer shape is determined by the effective fusion of the double layer regions, as caused by the relative distances and positions of the surfaces, clearly the topological shape of the double layers is inseparately linked with the geometry of the system. Indeed, if we put the trefoil knot of Figure \ref{fig:trefoil} on a smaller torus, we find a different set of topology transformations: instead of going from a $4-$torus to a $1-$torus, the shape actually changes to a $3$-torus instead, compare the second to third panel in Figure \ref{fig:tknots}(a) with Figure \ref{fig:trefoil}(e). The charge distribution $\sigma({\bf r})$ is, however, qualitatively the same, compare Figure \ref{fig:trefoil}(b) and the first panel in Figure \ref{fig:tknots}(a). 

To investigate how the interplay between the geometry and the topology of a knot affects the double layer topological shape, we consider the set of torus knots $T_{p,q}$, where $p$ is the number of turns the knot makes around the rotation axis of the ``virtual" torus on which the knot is positioned, while $q$ is the number of times the knot winds around an inner circle of the torus. For example, the trefoil knot $T_{3,2}$ is topologically equivalent to $T_{2,3}$ (which is also a trefoil knot). Although $T_{3,2}$ and $T_{2,3}$ are topologically equivalent, we see that the double layer behaves rather differently as function of $\lambda_D$. The double layer transforms from the topological shape of a trefoil knot to that of a 3-torus to a 1-torus to a sphere (see Figure \ref{fig:tknots}(b)), which is different than for $T_{3,2}$ (Figure \ref{fig:trefoil}(e) or Figure \ref{fig:tknots}(b)). This shows that also geometrical factors play a central role in determining the topology of $\Gamma_C$. 

To find a general trend for the topological transformations of $\Gamma_C$ for torus knots, we investigate a selection of torus knot particles with different $p$ and $q$ than the trefoil particle in Figure \ref{fig:tknots}(c)-(f), and characterise the topological shape of the resulting $\Gamma_{1/e}$. Knots can be topologically classified in various ways (including crossing number or tricolorability \cite{Adams:2004}), and in our case it is convenient to use the genus, which for a torus knot is given by $g=(p-1)(q-1)/2$. In this case $g$ is also a knot invariant, but is also defined for particles that are not knots (like the tori that we consider). For the isosurfaces that are not (torus) knots, $g$ is directly determined by counting the holes in the surface. We see that for torus knots there are two main routes for the topological shape transformations. We have the transformation 
\begin{equation}
\mathrm{knot}\rightarrow(p+1)\mathrm{-torus}\rightarrow p\mathrm{-torus}\rightarrow\mathrm{sphere}, \nonumber
\end{equation}
or
\begin{equation}
\mathrm{knot}\rightarrow(p+1)\mathrm{-torus}\rightarrow 1\mathrm{-torus}\rightarrow\mathrm{sphere}. \nonumber
\end{equation}
For the approximate value of $\lambda_D$ for which the topological shape of a sphere is attained for $\Gamma_{1/e}$, we plot $\sigma({\bf r})$ in the first column of Figure \ref{fig:tknots} with the value of $\sigma_\text{max}$ and $\sigma_\text{min}$ as function of $\lambda_D$ listed in the Supplementary material. 

We conclude that the topological shape of electric double layers for the geometry of knots shown here depends on where double layer overlaps upon increasing the Debye length $\lambda_D$, which occurs first at all the crossing points of the knot. For even larger $\lambda_D$, it depends on other geometrical parameters whether first the central hole of $\Gamma_C$ closes by increasing $\lambda_D$ or the outer $p$ holes. $\Gamma_C$ always form $p$ holes at the outside because these are the number of crossings of the particle with the inner hole of the torus where the particle is positioned on. Finally, we can continuously deform a torus knot in such a way that it is topologically equivalent to a torus knot, but that it is geometrically not a torus knot. In this case the topological transformation will also differ and the richness of the exact geometry of the particle can be broadly used to control where overlaps of the double layer will emerge.

\subsection*{\bf Non-trivial topological shape transformations of the double layer in Janus particles.} In the spherical, toroidal and torus knot particles that we have discussed, all topological shape transformations of the electric double layer could be rationalised by considering the particle and increasing its (tube) radius. From the resulting space curves, isosurfaces can be constructed with the exact same isosurface topology of the screening cloud. Here, we show, however, how properties of the double layer itself could result in non-trivial double layer topologies, by considering particles with a non-trivial surface patterning, such as Janus particles \cite{Glotzer:2007}. 

\begin{figure}[t]
\centering
\includegraphics[width=0.6\textwidth]{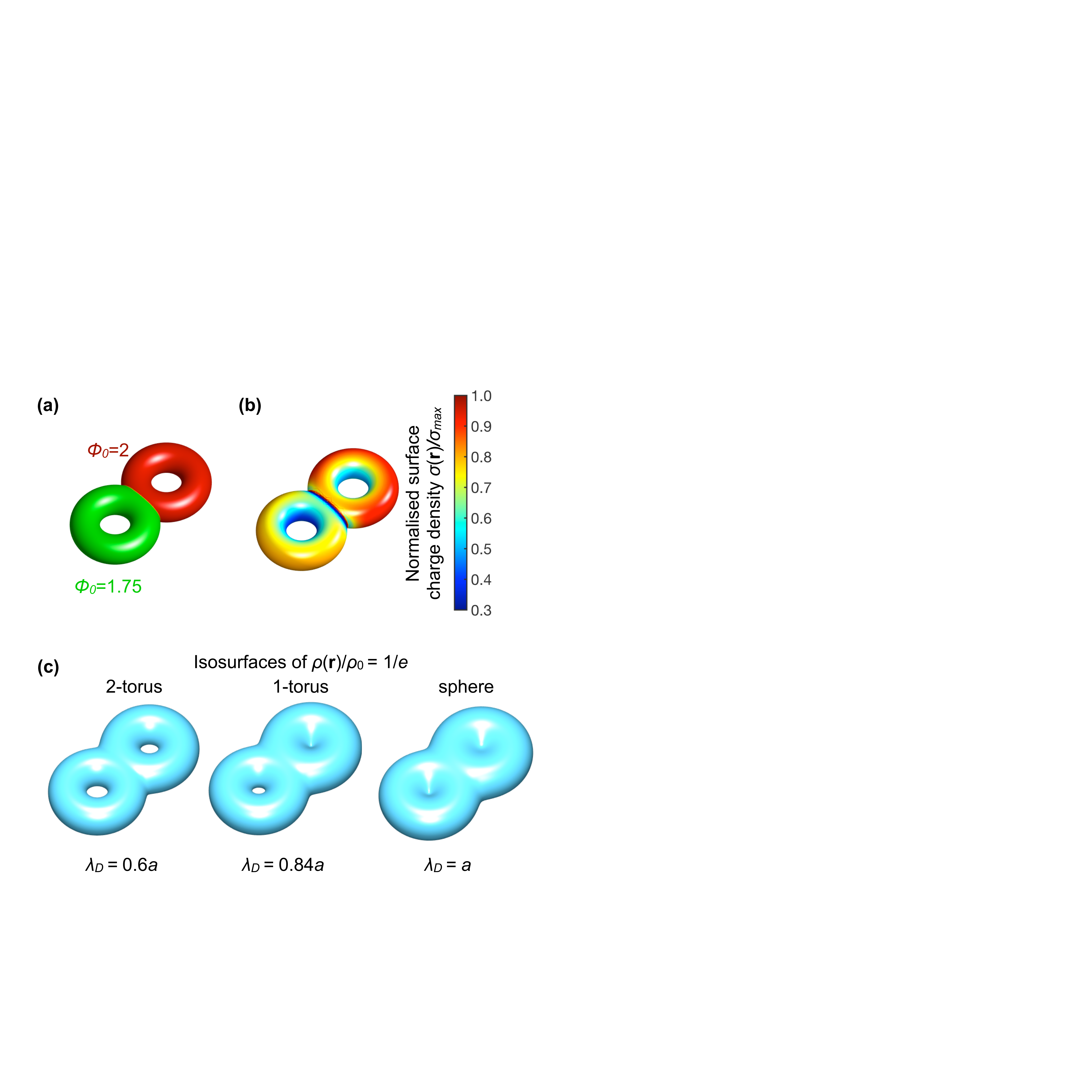}
\caption{\textbf{Double layer topology of a Janus double torus.} (a) Geometry of a Janus double torus, which is constructed by fusing two identical tori together with inner radius $a$, outer radius $2a$, separated by a center-to-center distance of $\Delta=2.4a$. One half has dimensionless surface potential $\Phi_0=1.75$, while the other half has $\Phi_0=2$. (b) Surface charge density $\sigma({\bf r})$ normalised on the maximal surface charge density $\sigma_\text{max}$ outside the interface region between the two tori. (c) Screening-cloud isosurfaces of constant net charge density $\rho({\bf r})/\rho_0=1/e$. These isosurfaces transform from a 2-torus topology via a toroidal topology to a spherical one by tuning the Debye length $\lambda_D$.}
\label{fig:dtorus}
\end{figure}

Firstly, we will consider the Janus double torus, which can be constructed by fusing two \emph{identical} tori of two different surface potentials together, see Figure \ref{fig:dtorus}(a). We choose both surface potentials to be positive, $\Phi_0>0$, which ensures the existence of screening-cloud isosurfaces that enclose the whole (positively charged) particle. Note that these isosurfaces would not exist for particles which have a positively charged and a negatively charged side. Such a fixed surface potential on a Janus particle produces again an inhomogenous surface charge distribution, as shown in Figure \ref{fig:dtorus}(b). The surface charge on the exact interface between the two different surface potentials are numerically obtained as noisy within the finite-element method, and are therefore not relevant for the analysis. For this reason, we normalised the surface charge density based on the $\sigma_\text{max}$ of the outer rims on the tori, sufficiently far away from this ``interfacial'' region. The global charging behaviour is what we would expect from what we have learned from the results of a single torus (Figure \ref{fig:torus}(b)-(c)), i.e., the highest charge density is found in the outer rim of the two tori. 
The double layer, however, transforms quite differently from what we would expect from the results in the previous section. We find the topological shape transformation 
\begin{equation}
\mathrm{2-torus}\rightarrow1\mathrm{-torus}\rightarrow\mathrm{sphere}. \nonumber
\end{equation}
The occurence of the 1-torus in this sequence cannot be rationalised from geometric arguments alone (the two holes in the particle have the same radius), and it is actually a direct result that the transition from one topology to another one depends on the surface potential of the surface. For the Janus double torus the hole of the torus of the \emph{higher} surface potential surface closes first, since more ions accumulate near the surface with the highest charge density therefore the screening clouds are fused together at a higher value of $\rho({\bf r})$ than at the side with the lower surface potential. 

However, this should not be confused with the results of the single tori, where we have seen that the decay of a lower surface potential surface is more pronounced with increasing $\lambda_D$: a torus of low surface potential closes its screening cloud hole at lower Debye lengths than a torus with high surface potential (Figure \ref{fig:torus}(e)), while in Figure \ref{fig:dtorus}(c) the hole with the high surface potential closes first. This difference in behaviour can be understood by noting that we normalised all the isosurfaces in the first row of Figure \ref{fig:torus}(e) by using $\Phi_0=1$, and in the second row we used $\Phi_0=2$. In Figure \ref{fig:dtorus}(c) we used $\Phi_0=2$ for the whole particle.

%The drawback is that the isosurface $\Gamma_{1/e}$ does not enclose the whole particle, since $\sinh(1)/\sinh(2)<1/e$. Instead we select, without loss of generality, the isosurface $\Gamma_{0.2}$. 
\begin{figure}[t]
\centering
\includegraphics[width=0.9\textwidth]{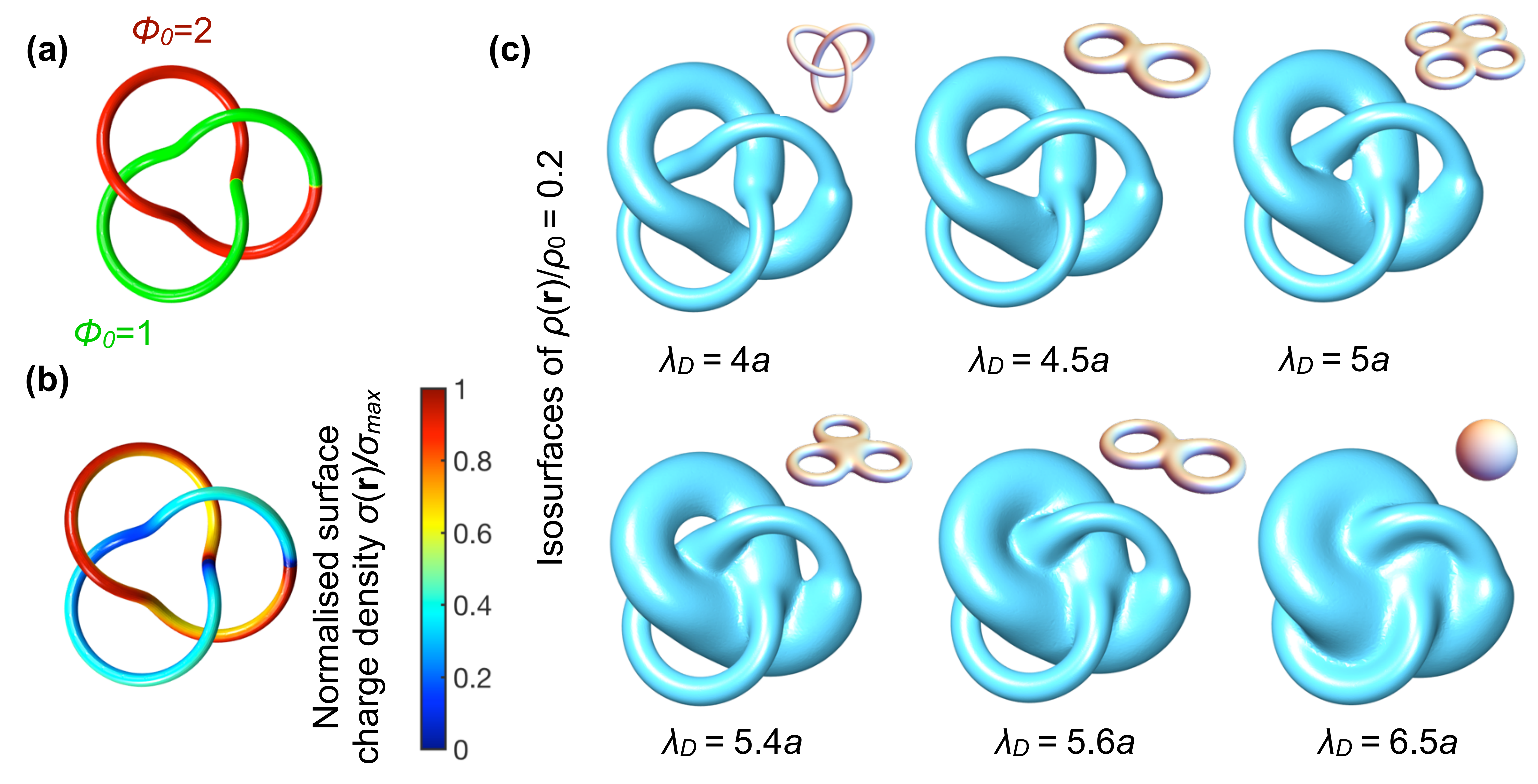}
\caption{\textbf{Topological transformation of the Janus trefoil knot double layer.} (a) Trefoil knot particle with fixed surface potential distribution $\Phi_0$. We use the same geometric parameters as in Figure \ref{fig:tknots}(a). (b) Inhomogeneous normalised surface charge distribution $\sigma({\bf r})/\sigma_\text{max}$ based on the fixed surface potential distribution. (c) Selected isosurfaces of constant charge density $\rho({\bf r})/\rho_0=0.2$ at various Debye lengths $\lambda_D$, with the equivalent topological shape in every upper-right inset.}
\label{fig:ptrefoil}
\end{figure}

Secondly, we consider the Janus trefoil knot particle as another example for a non-trivial transformation of the double layer topology. We construct this Janus particle from a tube with a fixed surface potential, connected to another tube with a different value for the surface potential, and subsequently fuse this patterned tube in a trefoil knot, see Figure \ref{fig:ptrefoil}(a). Same as with the double torus, this produces an inhomogenous surface charge distribution (see Figure \ref{fig:ptrefoil}(b)), where we used a $\sigma_\text{max}$ for normalisation that is well outside the interfacial region between the two surface potential regions, and we used a cut-off to visualise $\sigma({\bf r})$. To visualise the screening cloud topology, we use a surface potential of $\Phi_0=1$ and isosurface $\Gamma_{0.2}$ to produce non-trivial topological shapes of the screening cloud that could not be rationalised by geometric arguments of the particle alone, and to ensure that the selected isosurface encloses the whole particle. 
We find the following topological shape transformation (see Figure \ref{fig:ptrefoil}(c)),
\begin{equation}
\mathrm{knot}\rightarrow2\mathrm{-torus}\rightarrow 4\mathrm{-torus}\rightarrow 3\mathrm{-torus}\rightarrow 2\mathrm{-torus}\rightarrow\mathrm{sphere}. \nonumber
\end{equation}
The occurence of the 2-torus is the result of the interplay between the geometry of the particle with the distinct surface potential. Interestingly, the 2-torus occurs twice in this sequence. Moreover, the first 2-torus has an interesting geometrical shape, namely, the skeleton of this isosurface is actually the simplest singular knot that is produced by transforming an unknot to a trefoil knot by crossing the space curve of the knot with itself. This shows that the surface functionality of a particle can indeed have profound consequences on the resulting double layer topology with shapes that are also interesting from a purely mathematical perspective \cite{Adams:2004}.
\section*{Discussion}

This work explores the shape of double layers of complex-shaped particles in a simple environment with a special emphasis on the topological shape of the screening cloud. It is interesting to see how double layers of non-trivial topological shapes could be experimentally realised. We looked at particle shapes which currently can be synthesized using 3d printing, see for example Ref. \cite{Martinez:2014, Martinez:2015}. These particles are often rather large (typically in cca 10 micron regime as conditioned by the resolution of the printing method), therefore, not necessarily in the Brownian regime. A possible direct route to probe the double layer topology around such particles is by the use of fluorescent ions in combination with three-dimensional super-resolution fluorence microscopy techniques \cite{Leonhardt:2010}. Indeed, one could possibly explore the results of this paper by measuring specific isosurfaces of the mapped-out density profiles. However, to our best knowledge the double layer has not been measured in this way, not even for charged flat plates.

A second, but indirect{\mycolor, and more accessible} route to probe the complex double layers could be via measurements of the effective interaction between a pair of complex-shaped particles, e.g., via blinking optical tweezers \cite{Bartlett:2016}. Namely, the full shape of the double layer (i.e., the collection of all isosurfaces) directly affects the interaction potentials between particles, which could not only affect interaction of pairs, or clusters of particles, but also large-scale assembly of the colloidal disperson as a whole. {\mycolor In this case, we suggest an experiment where we start with the simplest non-trivial particle shape, i.e. a constant-charge or constant-potential torus, and measure the interaction potential for various center-of-mass distances and orientations. The torus-torus interaction potential can be, for example, compared to the interaction potentials of torus knots in a salt concentration regime where the torus knot's double layer is dominated by that of a toroidal shape. Currently, we are working on theory to support such experiments.}

Microscopic information on the double layer might{\mycolor also} affect the self-assembly properties of the system, for example, as was observed for spheres where the topology of the phase diagram (as function of particle density and Debye length) can be tuned by changing the charge-regulation mechanism \cite{Everts:2015}. Speculatively, the exact shape and topology of the double layer could lead to the occurence of a variety of ordered and disordered colloidal phases. {\mycolor The experimental observation of these systems should be supported by simulation data, which is, however, now scarce for complex-shaped \emph{charged} particles, however, performing simulations will be more accessible if effective pair interaction potentials are available.} Also interestingly, the topologically shaped double layers could show in transport properties of the colloidal dispersion, such as diffusion of colloids and on electrokinetic phenomena \cite{Oshima:2006}. 

Furthermore, the effect of multivalent ions (correlation effects) \cite{Podgornik:2013}, packing effects (concentrated electrolytes) \cite{Hartel:2015, Perkin:2017}, or fluctuation effects \cite{Fushiki:1992} on the double layer topology as defined in this paper can be also interesting for future work. Indeed it would be interesting to model the density profiles even more realistically than the Poisson-Boltzman theory that we have used here, which, however, gives \emph{at least} good qualitative results.

Finally, in this work we showed that electric double layers of complex topological shape can be realised and that they can be directly tuned by varying the Debye length, the exact geometry of the particle, and the surface functionality. Our findings are a possible step to introduce topological concepts to charged colloidal suspensions, which could lead to the design of new effective interactions and new ordered structures with interesting material properties.

%\begin{figure}[t]
%\centering
%\includegraphics[width=0.45\textwidth]{torus.pdf}
%\caption{Some particle shapes with nonSome particle shapes with nonSome particle shapes with nonSome particle shapes with nonSome particle shapes with nonSome particle shapes with nonSome particle shapes with nonSome particle shapes with nonSome particle shapes with nonSome particle shapes with nonSome particle shapes with nonSome particle shapes with nonSome particle shapes with nonSome particle shapes with nonSome particle shapes with nonSome particle shapes with nonSome particle shapes with nonSome particle shapes with nonSome particle shapes with nonSome particle shapes with nonSome particle shapes with nonSome particle shapes with nonSome particle shapes with nonSome particle shapes with nonSome particle shapes with nonSome particle shapes with nonSome particle shapes with nonSome particle shapes with nonSome particle shapes with nonSome particle shapes with non}
%\end{figure}

%\bibliographystyle{naturemag_noURL} % Tell bibtex which bibliography style to use
%\bibliography{literature1} % Tell bibtex which .bib file to use (this one is some example file in TexLive's file tree)

\section*{Methods}
\subsection*{\bf Numerical modelling.}
In this paper the key ingredient to study the electric double layers around charged (colloidal) particles is the (mean-field) Poisson-Boltzmann equation
\begin{equation}
\nabla^2\phi({\bf r})=\lambda_D^{-2}\sinh[\phi({\bf r})],
\label{eq:PB}
\end{equation}
where $k_BT\phi({\bf r})/q_e$ is the electrostatic potential, $q_e$ the elementary charge, and $k_BT$ is the thermal energy. Eq. \eqref{eq:PB} can be derived easily by combining the Poisson equation with Boltzmann distributions for the cation and anion density $\rho_\pm({\bf r})=\rho_s\exp[\mp\phi({\bf r})]$, with $\rho_s$ the bulk ion density. The width of the electric double layer is set by the Debye screening length $\lambda_D=(8\pi\lambda_B\rho_s)^{-1/2}$, with $\lambda_B$ the Bjerrum length. Inside the particle the Laplace equation is solved $\nabla^2\phi({\bf r})=0$, because of the absence of external charges in the interior of the particle. On the colloidal particle $\Gamma$, we assume constant-potential boundary conditions
\begin{equation}
\phi({\bf r})=\Phi_0, \quad {\bf r}\in\Gamma,
\label{eq:CP}
\end{equation}
with $k_BT\Phi_0/q_e$ a predetermined, constant value for the electrostatic potential on the colloidal surface. We choose these boundary conditions because of (i) its simplicity (no need to specify $\lambda_B$ and easy to define double layer topology, see main text) and (ii) it corresponds to a physical charge regulation mechanism in which the particle acquires its charge by the simultaneous adsorption of cations \emph{and} anions, see Ref. \cite{Everts:2015}. With other boundary conditions the results are qualitatively the same, but the surface potential is not constant along the particle surface and there is less tendency for discharging (see Ref. \cite{Everts:2015}), meaning that inhomogeneities will be less pronounced in the surface charge distribution profile.

We solve the closed set of Eq. \eqref{eq:PB} and \eqref{eq:CP} by using the finite-element method (COMSOL) in a simulation box that is chosen large enough such that the potential goes to zero far from the particle to ensure that the single-particle picture is valid. It is then straightforward to obtain the surface charge densities $q_e\sigma({\bf r})$ for a single particle by evaluating
\begin{equation}
{\bf n}\cdot\nabla\phi({\bf r})=-4\pi\lambda_B\sigma({\bf r}), \quad {\bf r}\in\Gamma, \label{eq:evalcharge}
\end{equation}
with $\bf n$ the outward-pointing unit surface normal. The net charge density $q_e\rho({\bf r})=q_e[\rho_+({\bf r})-\rho_-({\bf r})]$ in the diffuse ion cloud around the colloidal particle can be found by 
\begin{equation}
\rho({\bf r})=-2\rho_s\sinh[\phi({\bf r})].
\end{equation}
For convenience we define the net ion density on the particle surface $\rho_0=-2\rho_s\sinh(\Phi_0)$ which is constant due to our choice of boundary condition. Of course one can impose other charge-regulation mechanisms, but then both $\phi({\bf r})$ and $\sigma({\bf r})$ will vary along the particle surface, and these values will also depend on the thermodynamic state of the system. However, choosing this different boundary condition will not change the qualitative features of the single-particle results in this paper. In the supplementary information we show what would change if we would use constant-charge boundary conditions instead.

\subsection*{\bf Parametrization of torus knots.} The sphere and torus are parametrized in the standard ways. A $(p,q)$-torus knot with tube radius $a$ can be parametrized by ${\bf R}(u,v)=\boldsymbol{\gamma}(u)+{\bf n}(u)(a\cos v)+{\bf b}(u) (a\sin v)$, with $0\leq u,v\leq 2\pi$, where $\boldsymbol{\gamma}(u)$ is the parametric curve
  \begin{align}
{\boldsymbol{\gamma}}(u)= &= \left(\begin{matrix}
           [R+r\cos(qu)]\cos(pu) \\
          [R+r\cos(qu)]\sin(pu) \\
          r\sin(qu)
         \end{matrix}\right),
         \label{eq:tknot}
  \end{align}
and ${\bf n}(u)$ and ${\bf b}(u)$ are normal and binormal unit vectors of $\boldsymbol{\gamma}(u)$, respectively. Inspection of Eq. \eqref{eq:tknot} shows that a $(p,q)$-torus knot winds $p$ times around the axis of rotational symmetry of a torus with inner radius $r$ and outer radius $R$, and $q$ times around a circle in the interior of this torus.
The genus of a $(p,q)$-torus knot is given by $g=(p-1)(q-1)/2$, which is defined for a knot as the minimal genus of a Seifert surface of the knot (with a Seifert surface being a surface which has the knot, in this case $\boldsymbol{\gamma}(u)$, as a boundary).
\\
\\
\section*{\normalsize{Acknowledgements}}
J. C. E. acknowledges financial support from the European Union's Horizon 2020 programme under the Marie Skłodowska-Curie grant agreement No. 795377. M. R. acknowledges financial support from the Slovenian Research Agency ARRS under contracts P1-0099 and L1-8135. The authors acknowledge fruitful discussions with S. \v Copar and A. van Blaaderen.

\section*{\normalsize{Author contributions}}
J. C. E. performed the theoretical modelling and numerical calculations under supervision of M. R. Both authors co-initiated the topic and contributed to writing and discussing the manuscript.

\section*{\normalsize{Additional information}}
The authors declare no competing interests.

\end{document}